\documentclass[%
superscriptaddress,
groupedaddress,
preprint,
 amsmath,amssymb,
]{revtex4-1}

\usepackage{graphicx}
\usepackage{dcolumn}
\usepackage{bm}
\usepackage{amsmath}
\usepackage{epsfig}
\usepackage{rotating}
\usepackage{enumerate}
\usepackage{xcolor}
\usepackage{color,soul}
\usepackage{siunitx}
\usepackage{adjustbox}

\begin{document}

\title {Electronic transport modulation on few-layers suspended MoS$_2$ under strain}

\author{Igor Neri}
\affiliation{NiPS Laboratory, Dipartimento di Fisica e Geologia,
             Universit\`a degli Studi di Perugia, 
             06123 Perugia, Italy}
\email{igor.neri@nipslab.org, miquel.lopez@nipslab.org}
\author{Miquel L\'opez-Su\'arez}
\affiliation{NiPS Laboratory, Dipartimento di Fisica e Geologia,
             Universit\`a degli Studi di Perugia, 
             06123 Perugia, Italy}

\date{\today}

\begin{abstract}
The production of new sensors, transducers and electronic components can benefit from the possibility to alter the electronic transport of metal-semicondutor-metal (MSM) devices. 2D materials are extremely appealing for those new technologies. This can determined by several phenomena as piezoelectric effect, piezoresistive effect and modulation of Schottky barrier. In particular, MoS$_2$, among other Transition Metal Dichalcogenides (TMDs), is predicted to show a transition from semiconductor to metal under strain. 
In this article we present measurements on the modulation of electronic transport on few layer MoS$_2$ suspended ribbons under uniaxial tensile strain. Experimentally observed changes in the two terminal IV curves can be explained in terms of band gap closing in the semiconductor. A maximum gauge factor of 240 is achieved for a 3-layer ribbon. We also report on the fabrication process that allows to apply high strains to suspended MoS$_2$ ribbons, paving the way to future studies on the effect of strain in 2D materials.
\end{abstract}

\maketitle

\section{Introduction}
The study of the electronic properties of nanostructured materials has gained a lot of interest in the last years. Nanowires, thin films and nanoribbons have been considered as the base for applications as sensors \cite{hempel2012novel}, transducers \cite{boeva2016few}, energy harvesters \cite{fan2016flexible} and electronic components \cite{shen2014mesoporous}. 2D materials are foreseen to be particularly interesting for such applications because of their outstanding optical, mechanical and electrical properties along with very strong electro-mechanical and opto-mechanical coupling\cite{bertolazzi2011stretching,wang2012electronics,wu2014piezoelectricity}.
Graphene, hexagonal Boron Nitride (h-BN) and Transition Metal Dichalcogenides (TMDs) are examples of such materials.

Molybdenum disulphide, MoS$_2$, is a layered crystalline TMD with an hexagonal structure. Single layer MoS$_2$ is formed by a plane of Mo atoms sandwiched and covalently bonded to two planes of S atoms. Few layer and bulk MoS$_2$ is formed by successive stacking of this hexagonal structure through vdW forces. In contrast with graphene, MoS$_2$, along with the rest of TMDs and h-BN, is non centrosymmetric and has been theoryzed to show macroscopic polarization under mechanical strain \cite{duerloo2012intrinsic}. This property would be present (absent) for odd (even) layered ribbons. Recently, the piezoelectric response of MoS$_2$ has been experimentaly measured \cite{wu2014piezoelectricity,zhu2015observation} along with the first experimental insights on its piezoresistive response, indicating a sort of band gap modulation under mechanical deformation. MoS$_2$ has been suggested to show a transition from semiconductor to metal under mechanical strain\cite{scalise2012strain,johari2012tuning,ghorbani2013strain,conley2013bandgap} and not only for monolayer MoS$_2$ but also for few-layer MoS$_2$, and even for its bulk form\cite{yun2012thickness,lopez2016band}, which is partially supported by experimental data from Ref. \citenum{wu2014piezoelectricity}. In its bulk form, MoS$_2$ is a semiconductor with indirect band gap and Eg=\SI{1}{\electronvolt}, while its single layer counterpart shows a direct transition at the K point with Eg=\SI{1.9}{\electronvolt}. The band structure of MoS$_2$ is also influenced by the Schottky barrier that arise when in contact with a metal. The contribution of the Schottky barrier to the energy band-gap is expected to be of $0.06-0.16$ eV for Au-MoS$_2$ contacts\cite{kaushik2014schottky}.

In this work we report measurements of the change in electronic transport on few-layers suspended MoS$_2$ under tensile strain. Such behaviour can be accounted to the reduction of the energy band gap of the semiconductor and modulation of the Schottky barrier. Changes in the direct-current electronic transport are characterized for a set of MoS$_2$ suspended ribbons under uniaxial tensile strain. A drastic increment in the current flowing through the device is obtained when the material is effectively strained. The results and the methodology shown in this work pave the way for the development of micro and nanodevices exploiting the strong electro-mechanical coupling of MoS$_2$ such as strain sensors or straintronic devices, among others. In particular such strain induced semiconductor-conductor transition can be exploited to realize transistors where the gate is represented by the applied strain. In this prospective it is possible to imagine hybrid flexible computing devices that get the input from the human body movements (strain) producing an electrical output.

\begin{figure*}
\includegraphics[width=\textwidth]{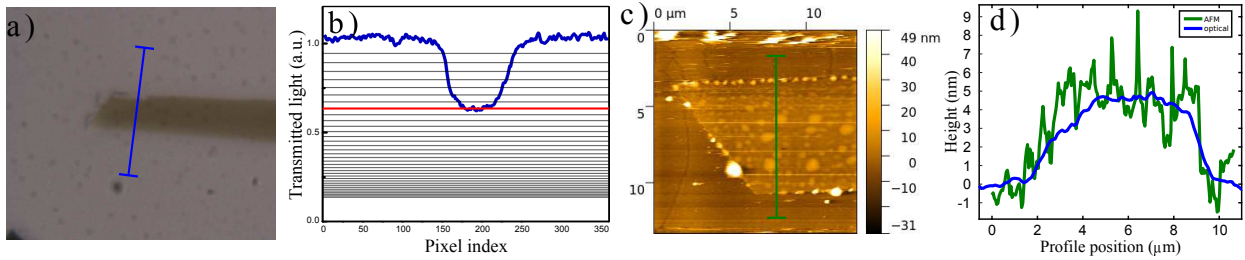}
\caption{Optical and AFM characterisation of a MoS$_2$ flake. a) Image under an optical microscope in light transmittance of a flake on PDMS. b) Estimation of the number of layers, $n$, based on the light transmittance of single layer MoS$_2$, corresponding to $n=8$. c) AFM measurement of the flake on the Au/Cr coated (\SI{55}{\nano\meter}) \SI{500}{\micro\meter} SiO$_2$ substrate. d) Height profile of the flake reconstructed from AFM image and light transmittance analysis.}
\label{f:layers}
\end{figure*}

\section{Methods}
MoS$_2$ flakes are obtained by mechanical exfoliation from a single crystal \cite{li2014preparation,li2012fabrication} with the ``scotch tape method'' and then transferred to a polydimethylsiloxane (PDMS) holder as in Ref. \citenum{castellanos2014deterministic}. An optical microscope is used to locate few layer MoS$_2$ candidates for the analysis. To estimate the number of layers we have performed an optical analysis considering the light transmittance of single-layer MoS$_2$ to be 94.5\% \cite{nl402875m}. The overall transmittance of the flake decreases with the number of layers following a power-law\cite{nl402875m}. An optical image of one of the flakes ($s_2$) is reported in Fig. \ref{f:layers} along with its transmittance profile and AFM measurement. Each horizontal gray line in the light transmittance profile is relative to the expected value for a specific number of layers, starting from the top line corresponding to a single layer, $n$=1. The red curve corresponds to the transmitted light for the estimated number of layers for the MoS$_2$ flake ($n=8$). The validity of the optical analysis is corroborated by AFM measurements reported on the third panel along with the estimated height from the optical analysis. The good agreement between the AFM measurement and the estimated height highlights the reliability of the layers estimation method from light transmittance. Optical images and analysis of the other flakes are reported in Supplementary Material.

\begin{figure}
\includegraphics[width=\columnwidth]{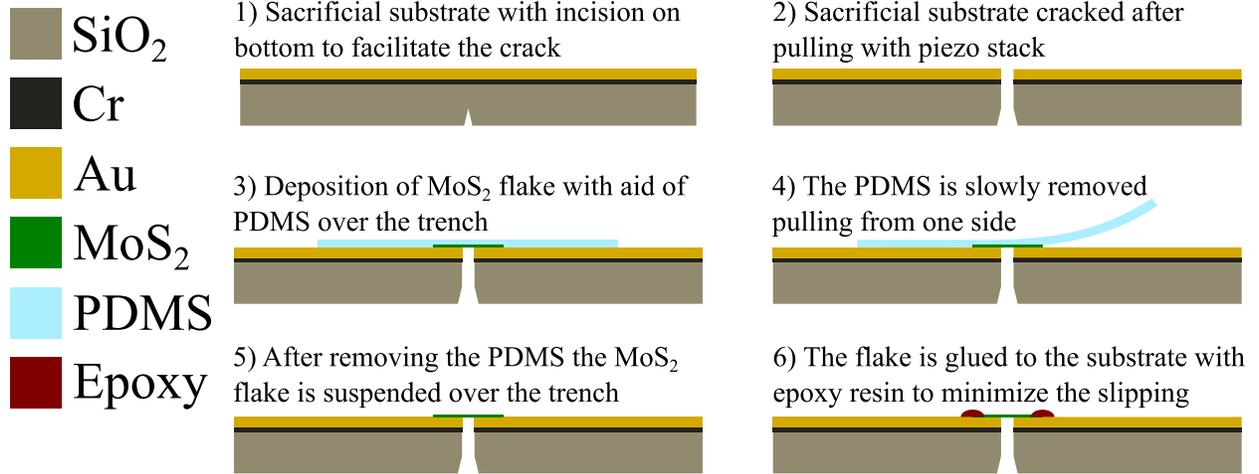}
\caption{Schematic representing the preparation of the hospital substrate and deposition of the MoS$_2$ flake.}
\label{f:scheme}
\end{figure}

The hospital substrate for the sample consists on a Au/Cr coated (\SI{55}{\nano\meter}) \SI{500}{\micro\meter} SiO$_2$ wafer. 
Before transferring the flakes each substrate has been mounted on a high precision nanopositioning stage clamping one end to the reference frame and the other to the movable part. The positioning system is capable of a maximum extension of \SI{35}{\micro\meter} with a pull force of \SI{10}{\newton}. An incision is made on the bottom of the substrate (see panel 1 in Fig. \ref{f:scheme}) in order to split it in two parts with a clean crack by means of the force applied by the piezoelectric stage. The two independent parts are then separated by a initial gap $g$ defining the two electrical contacts of the device. Selected flakes are transferred over the trench once the initial gap is set to $g=g_0$ where $g_0$ is the gap at which the current through the contacts, before transferring the flake, is less than \SI{0.01}{\nano\ampere}, negligible for the expected current passing through the device. This sets the length of the metal-semiconductor-metal (MSM) device to $l=g_0$.

The overall procedure for the preparation of the hospital substrate and deposition of the flake is presented in Fig. \ref{f:scheme} (see Supplementary Material for pictures taken during the flake transferring). It has to be noted that the application of strain to the suspended ribbon is a difficult task that can be influenced by different factors. In particular, the capacity to apply an effective strain depends on the adhesion forces between the flake and the substrate. If strain overcomes the adhesion forces the sample may slip over the contacts reducing drastically its internal strain. To tackle this problem it is possible to anchor the ends of the flake to the substrate (see panel 6 in Fig \ref{f:scheme}). However, this procedure may introduce additional strain even orthogonal to the gap opening direction that cannot be easily removed. For this reason we have conducted the experiments in both configurations, with and without anchoring.
Table \ref{t:samples} shows the main parameters of the devices. 

\begin{table}
\caption{Parameters of the measured samples.}
\begin{tabular}{|c|c|c|c|c|c|c|}
\hline 
Sample & $l (\mu m)$ & $w (\mu m)$ & $n$ & $A_L (\mu m^2)$ & $A_R (\mu m^2)$ & Clamped\\ 
\hline 
$s_1$ & 4.96 & 6.7 & 3 & 23.9 & 165.6 & no\\
\hline 
$s_2$ & 7.14 & 11.7 & 8 & 385.8 & 674.8 & no\\
\hline 
$s_3$ & 3.07 & 9.43 & 7 & 321.2 & 245.6 & no\\
\hline 
$s_4$ & 6.16 & 7.28 & 6 & - & - & yes\\
\hline 
$s_5$ & 9.2 & 2.92 & 7-bulk & - & - & yes\\
\hline 
\end{tabular} 
\label{t:samples}
\end{table}

To observe the modulation of the electronic transport as effect of applied strain we measured the current flowing through the material varying the polarization $V_\textnormal{p}$ in a two terminal configuration, \emph{i.e.} the IV curve. We recall here that applying tensile strain to MoS$_2$ results in a reduction of its energy band-gap up to the point of reaching a transition from semiconductor to metal (semi-metal for multilayer MoS$_2$). An additional collaterial effect of the strain can be the modulation of the Schottky barrier height\cite{quereda2017strain}.

Strain is applied to the MSM device by increasing the gap, $g$. The effective length of the gap is directly measured through the integrated positioning sensor of the linear actuator which has a resolution of \SI{0.2}{\nano\metre}. 

While in a digital simulation strain is uniformly applied along one or more dimensions of the device, in a real scenario small imperfections on the material itself, or the exact orientation in which the flake has been transferred to the host substrate can produce a non uniform distribution of strain. In order to know exactly the strain applied to the device a local measure with Raman and/or Photoluminescence spectroscopy should be considered, which is unfeasible for the current setup. In our experiments the strain is thus roughly estimated as $\epsilon=g/g_0-1$, where $g$ is the measured value of the gap, which bounds the effective maximum strain. For non clamped devices the estimated strain can be larger than the real one due to possible slippage of the sample over the substrate.

\begin{figure*}[!ht]
\includegraphics[width=\textwidth]{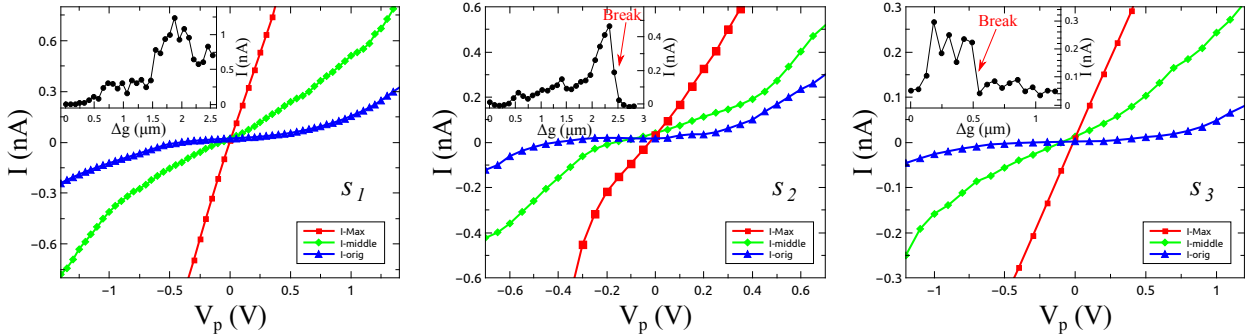}
\caption{Current-voltage relation as a function of applied tensile strain for $s_1$, $s_2$ and $s_3$. IV curves for the three unstrained devices are represented by blue triangles. The steepest IV curves, achieved at strains ranging from 15\% to 20\%, are represented by red squares. A linear relation is achieved for $s_1$ and $s_3$. Data corresponding to intermediate strains are showed with green circles. Insets show the current at fixed polarization voltage, $V_p=$\SI{300}{\milli\volt}, as gap increases. The rupture of the structure, highlighted with red arrows, corresponds to a drastic diminution of the current.}
\label{f:iv_free}
\end{figure*}

\section{Results and discussion}
We start our analysis with the not clamped devices. In Fig. \ref{f:iv_free} we show the two terminal IV curves for the three samples, $s_1$, $s_2$ and $s_3$, for three different strain levels: no strain, moderate strain, maximum strain. The typical semiconductor behaviour is observed for the pristine configuration (blue triangles): a linear relation in the IV curve at low bias voltage followed by a change in the slope at large biases.

Each inset in Fig. \ref{f:iv_free} shows the current flowing through the corresponding device at $V_p=$\SI{300}{\milli\volt} as $g$ is increased.
For the first sample it can be seen that, starting from $g=g_0$, the current increases as the gap is opened. In these first stretching steps the current reaches a value 2 orders of magnitude higher than this for $g=g_0$.
Samples $s_2$ and $s_3$ show similar qualitative trends but with a lower increase in the current of a factor 10 and 20, respectively.
The IV curves show that the increment of flowing current is accompanied by a linearisation of the IV relation, which can be explained in terms of energy gap closing although Schottky barrier height modulation cannot be excluded. Specifically, sample $s_1$ and sample $s_3$ show an increment in the slope of the IV curve from \SI{0.042}{\nano\ampere/\volt} to \SI{2.15}{\nano\ampere/\volt} and from \SI{0.0083}{\nano\ampere/\volt} to \SI{0.72}{\nano\ampere/\volt}, respectively. Sample $s_2$ equally shows an increment of more than two orders of magnitude, from \SI{0.0085}{\nano\ampere/\volt} to \SI{1.30}{\nano\ampere/\volt} yet preserving a non linear relation.
During strain application, MoS$_2$ flakes may slip over the Au contacts releasing part of the mechanical energy stored in the system recovering an electric response corresponding to a less strained state. Such events are visible on all insets of Fig. \ref{f:iv_free} as a drop on the current. 
The maximum strain that a device can support before slippage or rupture depends on several factors. Among them, the contact area between the flake and the substrate is expected to play an important role \cite{he2013experimental}. The areas of the left(right) contacts, $A_L$($A_R$), are reported on Table \ref{t:samples} for the non clamped samples. For sample $s_1$, $A_L$ is much smaller than $A_R$, and much smaller than the ones of the other samples. The reduced contact area favors potential slippage of the sample. As evident in the inset of panel (a) of Fig. \ref{f:iv_free}, during the last steps of the experiment, the current flowing through the device decreases, which may be caused by the slippage of the left contact. The presence of slippage is confirmed by analyzing the optical images of the devices taken during the experiment. The detected slippage can be correlated with the drop in the current corroborating the dependence of the increase in current flowing through the device with applied strain (see Supplementary Material for slippage analysis). This is accompanied also with a further reduction of the current due to a smaller contact area. Instead, sample $s_2$ and sample $s_3$ have much larger contact areas that helped preventing slippage, in this scenario during the application of the strain the flakes broke in the final steps as indicated with red arrows in Fig. \ref{f:iv_free}.

\begin{figure}
\includegraphics[width=0.8\columnwidth]{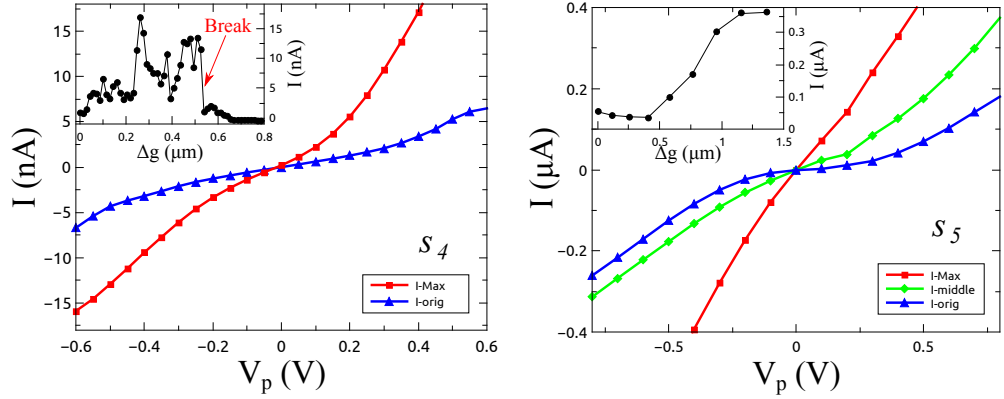}
\caption{Current-voltage relation as a function of applied tensile strain for $s_4$ and $s_5$. IV curves for the two unstrained devices are represented by blue triangles. The steepest IV curves are represented by red squares. A linear relation is achieved for $s_5$ while $s_4$ preserves a non linear response even if it shows a moderate current increase. Data corresponding to an intermediate strain are shown with green circles for $s_5$. Insets show the current at fixed polarization voltage, $V_p=$\SI{300}{\milli\volt}, as gap increases. A rise of the current is observed for both devices although for $s_4$ the increment is lesser than an order of magnitude. The rupture of $s_4$ is highlighted with a red arrow.}
\label{f:iv_clamped}
\end{figure}

In order to prevent slippage, we have repeated the measurements with two samples this time clamping both ends of the suspended flakes with epoxy resin, as depicted on panel 6 of Fig. \ref{f:scheme}. The same procedure for strain application is followed for these samples. The results are presented in Fig. \ref{f:iv_clamped}. For both samples the pristine semiconductor IV curve became steeper as strain is applied. However, sample $s_4$ still preserves a non linear IV relation even for the most strained states even if the IV curve is steeper in the central region reaching a maximum increment in the current at $V_p=$\SI{300}{\milli\volt} of a factor $2$. Sample $s_5$, composed by a bulk/7-layer heterostructure, presents more interesting features. In particular, the IV curve reached a linear relation for the maximum strain. Moreover, at fixed $V_p=$\SI{300}{\milli\volt}, the current is a monotonic function of the gap, suggesting that anchoring effectively prevents slippage over the contacts. The presence of the bulk material contributes to the mechanical stability of the few layer part of the ribbon, which reflects on less noisy data, while its contribute to the electronic transport can be neglected (see Supplementary Material for IV curve of the bulk part). 

\begin{figure}
\includegraphics[width=\columnwidth]{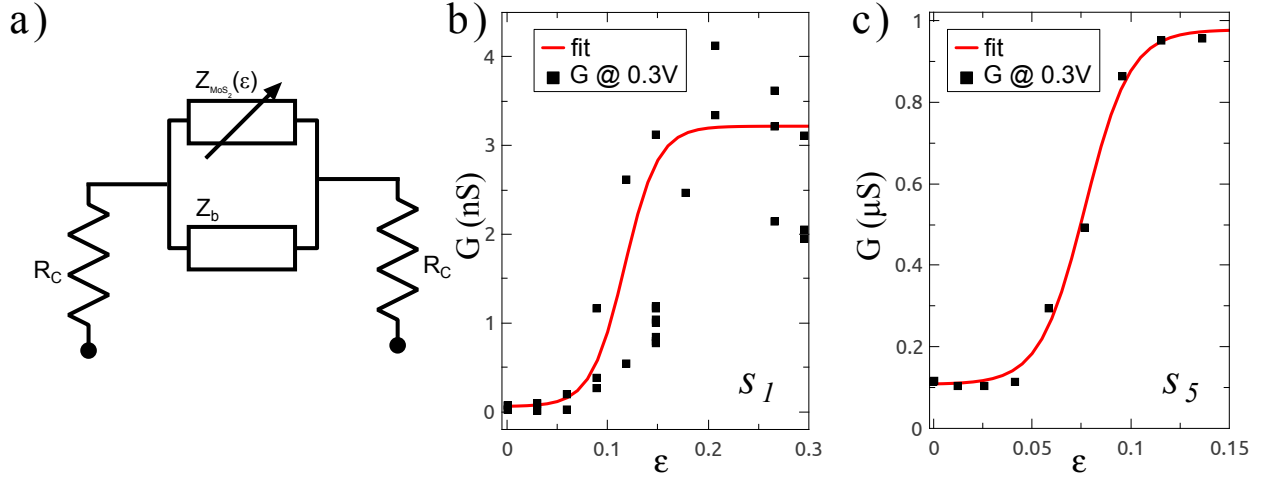}
\caption{Conductance of the device. a) Equivalent circuit of the device formed by two ohmic contacts in series with the strain dependent impedance of the MoS$_2$ flake and a background impedance. b) Measured (black squares) and modeled (solid red line) conductance of the device at $V_\textnormal{p}$=\SI{300}{\milli\volt}.}
\label{f:model}
\end{figure}

For samples $s_1$ and $s_5$ it is possible to estimate the applied strain as function of gap aperture and thus these samples give the opportunity to design and analyze the equivalent circuit of the measured device. A model of the equivalent circuit is presented in Fig. \ref{f:model}(a). $R_\textnormal{C}$ represents the contact resistances, $Z_\textnormal{b}$ the background impedance that accounts for the bulk MoS$_2$, for sample $s_5$, and parasitic currents. Finally, $Z_{\textnormal{MoS}_2}(\varepsilon)$ is the strain dependent impedance of the MoS$_2$ flake. According to the presented model the total conductance of the device can be expressed as:
\begin{equation}
G(\varepsilon)=\frac{G_\textnormal{R} \left(G_{\textnormal{MoS}_2} (\varepsilon)+ G_\textnormal{b}\right)}{G_\textnormal{R}+ G_{\textnormal{MoS}_2}(\varepsilon)+ G_\textnormal{b}}
\label{eq:1}
\end{equation}
where the conductance $G_x$ is the inverse of the impedance $Z_x$. Within this model the modulation of the Schottky barriers is neglected and only the MoS$_2$ conductance has a direct dependence with strain.
According to Ref. \citenum{manzeli2015piezoresistivity} the conductance of a semiconductor under strain can be expressed as
\begin{equation}
G_{\textnormal{MoS}_2}(\varepsilon)=G_0 \exp \left[-\frac{\varepsilon}{2 k_\textnormal{B} T }\frac{\partial E_\textnormal{g}}{\partial \varepsilon}\right]
\label{eq:2}
\end{equation}
where $G_0$ stands for the conductance of the unstrained MoS$_2$ flake, $k_\textnormal{B}$ is the Boltzmann constant and $T$ is the temperature.
The measured conductances of the devices are represented by black dots in Fig. \ref{f:model}(b) and (c) for sample $s_1$ and $s_5$, respectively. The model defined by Eq. \ref{eq:1} and Eq. \ref{eq:2} is used to fit experimental data showing a good agreement. The fitting parameters for the two samples are reported in Table \ref{t:parameters}, which agree with the literature\cite{manzeli2015piezoresistivity}. In particular the model gives an energy gap modulation of \SI{-31}{\milli\electronvolt\per\percent strain} for $s_1$ and \SI{-45}{\milli\electronvolt\per\percent strain} for $s_5$, close to the rate obtained from \textit{ab initio} calculations (\SI{-50}{\milli\electronvolt\per\percent strain})\cite{yun2012thickness,lopez2016band}.

\begin{table}
\caption{Fitted parameters from the conductance model for sample $s_1$ and $s_5$.}
\begin{tabular}{|c|c|c|}
\hline 
Parameter & $s_1$ & $s_5$ \\ 
\hline 
$R_\textnormal{C}$ & \SI{310.7e6}{\ohm} & \SI{1.02e6}{\ohm} \\
$Z_\textnormal{b}$ & \SI{1.67e10}{\ohm} & \SI{8.24e6}{\ohm} \\
$G_0$ & \SI{2.94e-12}{\siemens} & \SI{13.05e-10}{\siemens} \\
$\partial E_\textnormal{g} / \partial \varepsilon$ & \SI{-31}{\milli\electronvolt\per\percent strain} & \SI{-45}{\milli\electronvolt\per\percent strain} \\
\hline 
\end{tabular} 
\label{t:parameters}
\end{table}

We have calculated the gauge factor for the four MSM devices that achieved a linear IV relation ($s_1$, $s_2$, $s_3$ and $s_5$) at $V_\textnormal{p}$=\SI{300}{\milli\volt} ($\left[ \Delta I(\varepsilon) / I(0) \right] / \Delta \varepsilon $). For $s_1$ we have obtained $240$ in the range 0-20\% of strain. For $s_2$ and $s_3$ the corresponding gauge factors are $58$ and $71$ in the 0-15\% range, respectively. Finally for $s_5$ the gause factor is $118$ in the range 5-10\% of strain. The obtained values for all samples exceed the values of conventional and other 2D based strain sensors, like graphene based devices\cite{huang2011electronic}.

\section{Conclusions}

In conclusion, we have measured the changes in the electronic transport of few-layer MoS$_2$ ribbons under monoaxial tensile strain. An increase of more than an order of magnitude of the current flowing through the device and a linear IV relation are obtained for several devices at strains ranging from 10\% to 20\%. A maximum gauge factor of 240 has been obtained for a 3-layers MoS$_2$ ribbon, exceeding the values of other 2D based strain sensors. The observed increase in electronic transport is compatible with band-gap closure induced by strain, as suggested by previous \textit{ab initio} studies. The rate at which the band-gap is reduced, is estimated with a sound model of the conductance as function of the applied strain, obtaining data between $-31$ and \SI{-45}{\milli\electronvolt\per\percent strain}, close to the theoretical prediction. The results and the methodology shown in this work pave the way for the development of new devices at micro and nano-sale which exploit electro-mechanical coupling of MoS$_2$.

\section*{Author Contributions}
I.N. and M.L.S. contributed equally to this work.

\section*{Acknowledgments}
The authors gratefully acknowledge W. Venstra for useful discussion, S. Tacchi for AFM measurements. The authors acknowledge financial support from the European Commission (FPVII, Grant agreement no: 318287, LANDAUER).

%

\end{document}